\documentclass[a4paper,12pt]{article}
\usepackage[utf8x]{inputenc}
\usepackage{amsmath,amssymb,cite}
\usepackage{epsfig}
\usepackage{bbold}
\usepackage{hyperref}

\numberwithin{equation}{section}

\newcounter{appendice}

\title{Spontaneous Lorentz Violation in Gauge Theories}
\author{A. P. Balachandran$^{a,b,c}$\footnote{balachandran38@gmail.com}\, and S. Vaidya$^d$\footnote{vaidya@cts.iisc.ernet.in} \\
\begin{small}{\it $^a$Department of Physics, Syracuse University, Syracuse, N. Y. 13244-1130, USA}
\end{small}\\
\begin{small}{\it $^b$Institute of Mathematical Sciences, Chennai, India} \end{small} \\
\begin{small}{\it $^c$Instituto Internacional de Fisica, Natal, Brasil} \end{small} \\
\begin{small}{\it $^d$Centre for High Energy Physics, Indian Institute of Science, Bangalore, 560012, India}
\end{small}}

\date{\empty}

\begin{document}

\maketitle

\begin{abstract}
Frohlich, Morchio and Strocchi long ago proved that Lorentz invariance is spontaneously broken in 
QED because of infrared effects. We develop a simple model where the consequences of this breakdown 
can be explicitly and easily calculated. For this purpose, the superselected $U(1)$ charge group of QED 
is extended to a superselected ``Sky" group containing direction-dependent gauge transformations at infinity. 
It is the analog of the Spi group of gravity. As Lorentz transformations do not commute with Sky, they are 
spontaneously broken. These abelian considerations and model are extended to non-Abelian gauge 
symmetries. 

Basic issues regarding the observability of twisted non-Abelian gauge symmetries and of the 
asymptotic ADM symmetries of quantum gravity are raised.
\end{abstract}

\section{Introduction}

Some time ago, Frohlich, Morchio and Strocchi \cite{Frohlich:1978bf,Frohlich:1979uu} and Buchholz 
\cite{Buchholz:1986uj} proved that Lorentz invariance is spontaneously broken in QED because of infrared effects. It is broken for the same 
reason that the Higgs field spontaneously breaks gauge symmetry: the Lorentz group cannot be 
implemented on the representation space of local observables.

In this paper, we develop a model for the calculation of this violation in scattering processes. It is similar to 
the one we developed earlier for the induced electric dipole moment from QCD $\theta$-angle 
\cite{Balachandran:2012bn}. A $(2+1)$-dimensional version of this approach was developed recently in 
\cite{Balachandran:2012md}. It also overlaps with the work of Buchholz \cite{Buchholz:1986uj} and Buchholz and Fredenhagen 
\cite{Buchholz:1981fj}. The calculation itself will be presented elsewhere in order to keep the length of 
the present paper under control. An additional contribution in this work is the extension of the above 
considerations to non-Abelian gauge theories like QCD where too spontaneous Lorentz violation is seen.  

In Section 2, we review the theory of gauge invariance. It emerges from the basic contributions of 
Hanson, Regge and Teitelboim \cite{Hanson:1976cn} and has been explained before. There are different kinds 
of transformation groups in gauge theories. There are those implementing the Gauss law and acting as 
identity on quantum states. The QED and QCD (global) charges also arise from Gauss law, but they may 
not vanish on quantum state vectors. Rather they serve to define superselection sectors or inequivalent 
representations of the local algebra of observables. We recall this difference, as it plays a role in our 
further considerations. 

In QED, charged sources emit infrared photons. They create a cloud of electromagnetic field at "far" 
distances, in the "sky" of the source. There are clear similarities between this cloud and the CMB radiation 
of cosmology. The charge operator does not reflect the possibility of this cloud which can carry net zero 
electric charge. 

The proper description of this cloud involves direction-dependent gauge transformations at large distances. 
The enlarged gauge group which includes such transformations is our Sky group. It is similar to the spatial 
infinity or Spi group in gravity literature \cite{Ashtekar:1978zz}. The Sky group of QED serves to define 
superselection sectors. This abelian Sky group is introduced in Section 3. 

Section 4 examines Lorentz symmetries in the presence of the Sky group. Unlike $U(1)$, generic elements 
of the Sky group are not Lorentz invariant, and therefore, we argue that in this sector there is spontaneous 
Lorentz symmetry breaking. 

Summing up, our understanding is this: the treatment of infrared photons requires angle-dependent 
gauge transformations at infinity and the latter break Lorentz symmetry. 

A concrete and simple model for the calculation of this breakdown in scattering processes does not exist in 
the literature. For this purpose in section 5 we introduce an algebra generated by two unitaries $U(\chi)$ and 
$V(\omega)$. The $\chi$ is a function on $\mathbb{R}^3$ with a limit $\chi_\infty$ as the radius 
$r \rightarrow \infty$:
\begin{equation}
\lim_{r \rightarrow \infty}  \chi(r, \hat{n}) = \chi_\infty (\hat{n}), \quad \hat{n} \cdot \hat{n} = 1. 
\end{equation}
The $\omega$ is a closed two-form, also with a limit $\omega_\infty$ as $r \rightarrow \infty$:
\begin{equation}
\lim_{r \rightarrow \infty} \omega (r,\hat{n}) = \omega_\infty(\hat{n}).
\label{omegalimit}
\end{equation}
They generate a Weyl-like algebra
\begin{equation}
U(\chi) V(\omega) = C(\chi_\infty, \omega_\infty) V(\omega) U(\chi), \quad C(\chi_\infty,\omega_\infty) =
{\rm central\,\,elements}.
\label{sky_algebra}
\end{equation}
The meaning of these operators is this: $U(\chi)$ represent the elements of the Sky group while $V(\omega)$ 
intertwines its different representations. 

The algebra (\ref{sky_algebra}) is only a modification of a Buchholz-Fredenhagen algebra 
\cite{Buchholz:1981fj} introduced by them also to discuss gauge theories. Local observables cannot affect 
$\chi_\infty$ or $C$ as they depend on data at infinity. We see in section 5 that the Lorentz group does 
change it, and so gets spontaneously broken. 

If $|\cdot\rangle_{C=0}$ is a vector state on which $U(\chi)$ is $\mathbb{1}$, then
\begin{equation}
V(\omega) |\cdot \rangle_0 = |\cdot \rangle_C
\label{vertex}
\end{equation}
is a vector state on which $U(\chi)$ has value $C$. We show in Section 5 how to see the effect of $C$ 
in scattering processes by a simple modification of the fermion mass. We note that we do not prescribe a 
choice for $C$ here. It is determined by the details of the scattering process and perhaps can be modeled from 
earlier work on the infrared problem. We hope to address this issue in our follow-up paper \cite{bqv}. 

There is an alternative choice for $V(\omega)$, call it $W(\theta)$, which commutes with the Gauss law 
and intertwines different eigenspaces of the (global) charge operator. It is
\begin{equation}
W(\theta) = e^{i \theta S_{CS}(A)}
\end{equation}
where $S_{CS}(A)$ is the abelian Chern-Simons action for the spatial slice of QED. For the fermion mass 
term modified as previously, the vector states 
\begin{equation}
|\cdot \rangle'_\theta = W(\theta) |\cdot \rangle'_{\theta=0}
\end{equation}
also lead to Lorentz symmetry violation (The prime is to distinguish these vectors from those in 
(\ref{vertex})). But the algebra with relation (\ref{sky_algebra}) is modified.

The twist $V(\omega)$ reverses to $V(-\omega) = V(\omega)^{-1}$ under ${\cal C}$ and ${\cal P}$ and is 
${\cal T}$-invariant, while $W(\theta)$ is ${\cal C}$-invariant while reversing under ${\cal P}$ and 
${\cal T}$. Thus $|\cdot \rangle_\chi$ for $\chi \neq 1$ breaks ${\cal C}$ and ${\cal P}$ while 
$|\cdot \rangle_\theta$ for $\theta \neq 0$ breaks ${\cal P}$ and ${\cal T}$. Both keep ${\cal CPT}$ intact.

Section 6 takes up non-Abelian gauge groups. The analog of $U(\chi)$ exists here as well and serves to 
define the non-Abelian Sky group. But a $V(\omega)$ commuting with Gauss law does not exist. Instead, 
we must perforce use the non-Abelian $W(\theta)$ where $S_{CS}(A)$ is the non-Abelian Chern-Simons 
action for the spatial three-manifold. We get a new algebra from $U(\chi)$ and $W(\theta)$ and a new model 
for Lorentz violation.

With Section 7, we close with qualitative remarks on {\it non-Abelian} gauge invariance. Thus for example 
observables commute with {\it all} gauge transformations. For this reason, if the gauge bundle is twisted such 
as in QCD, it immediately confronts us with issues like the "problem of color" 
\cite{Nelson:1983bu,Balachandran:1983xz, Balachandran:1983fg,Nelson:1983fn} and mixed states 
\cite{Balachandran:2011bv}.  
In quantum gravity where diffeos assume the role of the gauge group, and observables commute with them, 
asymptotic symmetries, including the Poincar\'e group cease to be observable.Only their center, like mass 
and total spin can be measured. Thus for a quantum black hole, it is impossible to measure the probability 
of finding the component of angular momentum in a generic direction in any exact theory of quantum 
gravity. We explain these points in conclusion. 

\section{QED: The Gauss Law and the Charge}

In the canonical approach to QED, the equal-time vector potentials $A_i$ and the electric fields $E_i$ are 
canonically conjugate:
\begin{equation}
[A_i(\vec{x},t), E_j(\vec{y},t)] = i \delta_{ij} \delta^3(x-y).
\end{equation}
In elementary treatments, the classical Gauss law is imposed as a constraint on quantum vector states 
$|\cdot\rangle$:
\begin{equation}
(\partial_i E_i + j_0) |\cdot \rangle =0, \quad j_0 = {\rm charge\,\,density}.
\end{equation}

It has been understood for a long time that this formulation of quantum Gauss law is unsound: $A_i$ and 
$E_j$ are operator-valued distributions so that their derivatives must be reformulated in terms of test 
functions, their duals. For test functions, we can pick $C_0^\infty(\mathbb{R}^3)$, which is the space of 
infinitely differentiable functions on $\mathbb{R}^3$ vanishing at $\infty$. Note that we do not require 
the derivatives of functions to vanish at infinity.

Then the quantum Gauss law is
\begin{equation}
G(\Lambda) |\cdot \rangle = \int d^3 x [-(\partial_i \Lambda) E_i + \Lambda j_0](\vec{x},t) |\cdot \rangle =0, 
\quad \forall \Lambda \in C_0^\infty(\mathbb{R}^3).
\label{qGausslaw}
\end{equation}

For classical fields, we can here partially integrate without generating boundary terms, getting
\begin{equation}
\int d^3 x \Lambda(x)(\partial_i E_i + j_0)(\vec{x},t) = 0, \quad \forall \Lambda \in C_0^\infty(\mathbb{R}^3),
\end{equation}
which can be interpreted as the classical Gauss law.

Let us now enlarge $C_0^\infty (\mathbb{R}^3)$ to a larger space $C^\infty(\mathbb{R}^3;S^2_\infty)$. It 
consists of $C^\infty$-functions which have an angle-dependent limit $\chi_\infty$ as the radius variable $r$ 
goes to infinity:
\begin{equation}
\chi \in C^\infty(\mathbb{R}^3;S^2_\infty) \implies \lim_{r \rightarrow \infty} \chi(\vec{x}) = \chi_\infty (\hat{n}), \quad \hat{n} \cdot \hat{n} =1.
\end{equation}
It is possible to interpret this space as the space of functions after blowing up infinity, as in the treatment of 
Spi group. We are in effect attaching a sphere $S^2_\infty$ at infinity, with its points $\hat{n}$.

Clearly,
\begin{equation}
C^\infty_0(\mathbb{R}^3) \subset C^\infty(\mathbb{R}^3;S^2_\infty)
\end{equation}

We can now also consider the operators
\begin{equation}
Q(\chi) = \int d^3x \big(-\partial_i \chi E_i  + \chi j_0 \big)(\vec{x},t )
\label{charge}
\end{equation}
There is no reason for $Q(\chi)$ to vanish on physical states unless $\chi \in C^\infty_0(\mathbb{R}^3)$. In that 
case $Q(\chi)$ becomes $G(\chi)$.


In quantum theory, it is $Q(\chi)$ which gives the charge operator $Q_0$ after the condition 
$\chi_\infty (\hat{n}) =1$: 
\begin{equation}
Q_0 = Q(\chi)|_{\chi_\infty(\hat{n})=1}.
\end{equation}
The global $U(1)$ group is generated by $Q_0$. 

\section{The Sky Group}
We generalize the global $U(1)$ group to a bigger "Sky" group, call it Sky, with generators $Q(\chi)$. Of 
course only its quotient by Gauss law group can act nontrivially on the quantum states. So what we mean 
by Sky is the quotient of the group generated by $Q(\chi)$ by the Gauss law group. \\

\noindent {\it The Need for Sky} \\

In QED, infrared photons of arbitrarily small wavelengths are emitted by the sources. Even if at some time 
$t=-T$, we assume that there are no such photons for $r>R$, and that charge densities are compactly 
supported, the photons they emit will reach the $t=0$ surface for $r>R$ if $T$ is large enough (see Figure 
1). So we cannot rule out radiation field at arbitrarily large distances.  \\

\begin{figure}[htb]
		\begin{center}
\includegraphics[height=9cm]{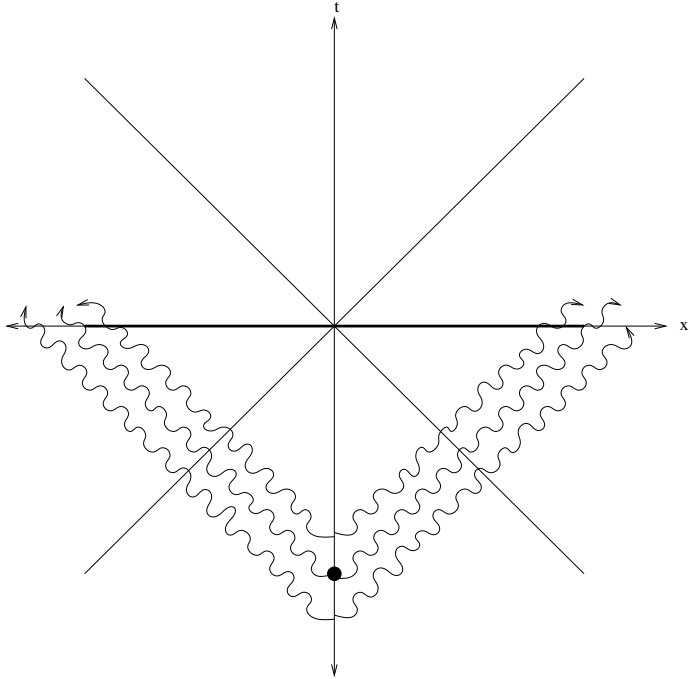}
		\caption{Photons from the distant past arrive at the $t=0$ surface at large distances.}
		\end{center}
		\end{figure}

But then, for their proper description using $A_i$ and to treat gauge invariance, we should allow 
gauge transformations at arbitrarily large distances. Neither the Gauss law $G(\Lambda)$ nor the charge 
$Q_0$ generate these transformations. So we expand global $U(1)$ to Sky. 

We work in the $A_0=0$ gauge. For this reason, we need not introduce time dependence in the elements of 
the Sky group. \\

\noindent {\it Sky is superselected}\\

{\it All} observables must preserve the physical state condition (\ref{qGausslaw}). If ${\cal A}$ is the algebra of observables and $\alpha \in {\cal A}$, we thus have
\begin{equation}
[G(\Lambda), \alpha]=0.
\end{equation}
We require that ${\cal A}$ is local as well. A good way to explain locality in this context is as follows. In 
a Lagrangian quantum field theory, fields $\phi$ and their conjugate momenta $\pi$ all commute at distinct 
spatial points at equal times:
\begin{equation}
[\phi(\vec{x},t),\pi(\vec{y},t)]=0.
\end{equation}
We can define local operators as their spatial integrals at a fixed time with test functions of compact spatial 
support. 

Locality implies that Sky commutes with ${\cal A}$: it lies in the commutant ${\cal A}'$ of ${\cal A}$:
\begin{equation*}
{\rm Sky} \subseteq {\cal A'}
\end{equation*}
To see this, let $\alpha \in {\cal A}$ be localized on a compact set $K$. Now for every $\chi$, we can find a 
$\Lambda$ such that $\chi$ and $\Lambda$ coincide on $K$:
\begin{equation}
\chi \restriction K = \Lambda \restriction K
\end{equation}
which means that
\begin{equation}
Q(\chi \restriction K) = G(\Lambda \restriction K).
\label{QisG}
\end{equation}
But since $\alpha$ is localized in $K$, by definition,
\begin{eqnarray}
[Q(\chi), \alpha] &=& [Q(\chi \restriction K), \alpha], \\ 
\protect{[}G(\Lambda), \alpha] &=& [G(\Lambda \restriction K), \alpha].
\end{eqnarray}
Using (\ref{QisG}), we then get
\begin{equation}
[Q(\chi), \alpha] = 0.
\end{equation}
This means that $Q(\chi)$ are superselected. They play the role of Casimir operators for Lie group 
representations. Their distinct eigenvalues label inequivalent representations of local algebras of observables.

\section{The Poincar\'e Group as Global Symmetry}

The generators of Poincar\'e group, just as those of any global symmetry, involve integrations of local fields 
over all space. They do not commute with {\it all} local fields localized in any compact spatial region. But as 
the group they generate preserves the algebra of local observables, they form an automorphism of this 
algebra. But such an automorphism may or may not be implementable in a particular representation of 
this algebra. 

A simple example can illustrate this point. The group $SU(3)$ and its group algebra $\mathbb{C}SU(3)$ 
admit complex conjugation or charge conjugation as its automorphism. Now $SU(3)$ (and therefore 
$\mathbb{C} SU(3)$) has two inequivalent three-dimensional irreducible representations (IRRs), the {\bf 3} and 
the ${\bf \bar{3}}$. The above automorphism exchanges them. If we consider just the IRR {\bf 3}, then 
the above automorphism cannot be implemented on the representation space.

Now going back to the charge operator $Q_0$, it is Poincar\'e invariant because current is conserved. The 
proof relies on the fact that if $g$ is a Poincar\'e transformation and $g^*\chi$ is the Lorentz transform of 
the test function $\chi$ with constant value $\tilde{\chi}$ at infinity, then $g^* \chi$ has the same value 
$\tilde{\chi}$ at infinity.

We conclude that the Poincar\'e group respects charge superselection rule. The latter places no obstruction 
to its unitary implementation.

For theories with mass gap, there is no further obstruction. The Poincar\'e group in such theories is not 
spontaneously broken. 

But for the full Sky group, the conclusions are different. If $\chi_\infty$ has non-trivial $\hat{n}$ 
dependence, and $R$ is a rotation, then $(R^* \chi)_\infty (\hat{n}) = \chi_\infty(R^{-1} \hat{n}) \neq 
\chi_\infty(\hat{n})$. Hence rotations, and therefore Lorentz transformations do not commute with those of 
Sky.

Spacetime translations in contrast are Sky-invariant. For example for spatial translations, this follows from 
the fact that
\begin{equation}
\frac{(\vec{x}+\vec{a})}{|\vec{x} + \vec{a}|} \stackrel{r \rightarrow \infty}{\longrightarrow} \frac{\vec{x}}{|\vec{x}|} 
= \hat{n}.
\end{equation}
As for time translations, since the Hamiltonian density 
\begin{equation}
{\cal H}(x) = \frac{1}{2} \Big( F_{0i}^2 + \frac{1}{2} F_{ij}^2 \Big) + \cdots
\end{equation}
itself is gauge invariant and commutes with $Q(\chi)$, elements of the Sky group also are 
time-translation invariant.

We conclude that unless
\begin{equation}
Q(\chi) |\cdot \rangle =0
\label{g_invariant}
\end{equation}
for $\chi_\infty$ not constant on $S^2_\infty$, the Lorentz group is spontaneously broken.

\section{The Sky Changes the $S$-matrix}
Let us start with quantum states $|\cdot \rangle_0$ carrying the trivial representation of Sky where 
(\ref{g_invariant}) is fulfilled. 

We will now construct the operator $V(\omega)$ of Section 1 which acting on $|\cdot\rangle_0$ creates 
the vector state 
\begin{equation}
|\cdot \rangle_\omega = V(\omega) |\cdot \rangle_0
\end{equation}
which carries a non-trivial Sky-representation, while at the same time it commutes with $G(\Lambda)$ and 
respects Gauss law.

With $V(\omega)$ at our disposal, there is a simple modification of the charged fermion-mass term which 
picks up the effect of $\omega$ and gives a Lorentz-violating $S$-matrix. Gauss law and charge conservation 
are maintained by the new interaction, while if $\omega=0$ and there is empty Sky, it does not affect 
scattering either. 

Let $\omega$ be a closed two-form,
\begin{equation}
d \omega = 0,
\end{equation}
and consider
\begin{equation}
V(\omega) = \exp i \left(Q_0 \int \omega \wedge A \right).
\label{Vdef}
\end{equation}
In the above, $Q_0$ commutes with $\int \omega \wedge A$. That is because on ${\mathbb R}^N$, any 
closed form is exact, and hence the integral of $\omega$ over any two-sphere is 0. So there is no 
ordering ambiguity in (\ref{Vdef}). The presence of $Q_0$ in $V(\omega)$ ensures that in any sector with 
net zero charge, $V(\omega)$ is identity and hence there is no Lorentz violation \cite{Frohlich:1978bf}.

$G(\Lambda)$ commutes with $V(\omega)$ since
\begin{equation}
\int \omega \wedge d \Lambda = -\int d \omega \wedge \Lambda = 0 
\end{equation}
for $\Lambda \in C_0^\infty$. Hence if $|\cdot \rangle$ is annihilated by $G(\Lambda)$, then so is 
$V(\omega |\cdot \rangle$:
\begin{equation}
G(\Lambda) V(\omega) |\cdot \rangle = 0. 
\end{equation}
This means that $V(\omega)$ maps admissible vectors compatible with Gauss law to other such admissible 
vectors. 

But if
\begin{equation}
U(\chi) = e^{i Q(\chi)} \in {\rm Sky}
\end{equation}
we see that 
\begin{equation}
U(\chi) V(\omega) = e^{-i Q_0 \int d \chi \wedge \omega} V(\omega) U(\chi) = C(\chi_\infty, \omega_\infty) 
V(\omega) U(\chi),
\end{equation}
where $C$ depends only on $\chi_\infty$ and $\omega_\infty$ since the integral $\int \omega \wedge 
d \chi$ depends only on them. 

This generalized ``Weyl" algebra is very similar to the Buchholz-Fredenhagen algebra 
\cite{Buchholz:1981fj}. Hence $V(\omega)$ maps an empty Sky $|\cdot\rangle_0$ to a non-empty one 
$|\cdot \rangle_\omega$.

The operator $V(\omega)$ is not a local observable, nor can it be identified as an observable in some sense, 
as it does not commute with Sky. It is like a charged field, but it commutes with $Q_0$. \\

\noindent {\it Remark}\\

$V(\omega)$ does not create charge as follows from (\ref{omegalimit},\ref{Vdef})
\begin{equation}
Q_0 V(\omega) |\cdot \rangle = V(\omega) Q_0 |\cdot \rangle
\end{equation}

We are after a simple model to calculate the effects of Lorentz breaking. There is in fact a simple approach 
to such a model. It was devised earlier \cite{Balachandran:2012bn} to calculate electric dipole moment 
from QCD $\theta$ and can be adapted for the present purpose.

Such a model is one where the fermion mass term ${\cal L}^0_m:= m \bar{\Psi}\Psi$ is modified to 
\begin{equation}
{\cal L}^\chi_m = \frac{m}{2} \big(U(\chi) + U(\chi)^\dagger\big) \bar{\Psi}\Psi.
\end{equation}
Since 
\begin{equation}
_\omega \langle \cdot | U(\chi) |\cdot \rangle_\omega = C(\chi_\infty, \omega_\infty) _0\langle \cdot | U(\chi) 
|\cdot \rangle_0 = C(\chi_\infty, \omega_\infty)\,\, _0\langle \cdot |\cdot \rangle_0,
\end{equation}
the amplitudes for the vectors $|\cdot \rangle_\omega$ are obtained from the amplitudes for the vectors 
$|\cdot \rangle_0$ if we replace $m$ by 
\begin{equation}
M(C(\chi_\infty,\omega_\infty)) \equiv m \Re C(\chi_\infty, \omega_\infty) 
= m \cos \Big( Q_0 \int d \chi \wedge \omega\Big). 
\end{equation}
We can then set $U(\chi)=1$, which is its value on $|\cdot \rangle_0$.

The Sky-deformed mass $M(C(\chi_\infty, \omega_\infty))$ depends in general on $\hat{n}$. Because of 
this dependence, it can give scattering amplitudes that violate Lorentz invariance. Its choice has to be 
dictated by the infrared cloud of a specific process. For instance, $\chi$ could be chosen to be the profile of 
the coherent state of soft radiation that accompanies incoming or outgoing charged particles. The explicit 
form of such is profile has been derived in \cite{Eriksson:1970dc}.\\

\noindent {\it An Alternative Choice for $V(\omega)$}\\

Instead of $V(\omega)$, we could have chosen any other unitary operator commuting with $G(\Lambda)$, 
but not commuting with a general $Q(\chi)$.

There is one particularly interesting choice of this sort. It is the one which generalizes to the non-Abelian case. 
It is
\begin{equation}
W(\theta) = e^{i \theta S_{CS}(A)}
\end{equation}
where
\begin{equation}
S_{CS} = \frac{1}{8\pi^2}\int A \wedge d A
\end{equation}
is the Chern-Simons term.

Then
\begin{equation}
U(\chi) W(\theta) = e^{-\frac{i \theta}{8 \pi^2} \int d \chi \wedge dA} W(\theta) U(\chi) := C(\chi_\infty, \theta) 
W(\theta) U(\chi).
\end{equation}
As before we can consider
\begin{eqnarray}
W(\theta) |\cdot \rangle_0  &=& |\cdot \rangle_\theta, \\
U(\chi) |\cdot \rangle_0 &=& |\cdot \rangle_0.
\end{eqnarray}
The mass term ${\cal L}^0_m$ between the $\theta$-sectors is equivalent to the mass term 
\begin{equation}
{\cal L}^\theta_m = M(\theta) \bar{\Psi} \Psi, \quad M(\theta) = m \cos \Big(\frac{\theta}{8\pi^2} \int d \chi 
\wedge dA \Big).
\label{csmass}
\end{equation}
in the $\theta=0$ sector. Since $dF=0$, this too depends only on asymptotic data. For the $A$ which occurs 
in (\ref{csmass}) as well, such data must come from the state of the infrared photons. 

\section{Non-Abelian Generalization}

The Gauss law and $Q(\chi)$ easily generalize to non-Abelian semi-simple compact Lie groups $H$.  It 
is enough to replace $\Lambda$ and $\chi$ by functions valued in its Lie algebra $\underline{H}$ and take 
traces at appropriate spots. Thus for instance
\begin{equation}
Q(\chi) = \int d^3 x \Big(- {\rm Tr} (D_i \chi) E_i + \chi J_0 \Big), \quad \chi = \chi^\alpha \lambda_\alpha, 
\quad \chi^\alpha \in C^\infty(\mathbb{R}^3;S^2_\infty).
\end{equation}
Here $\lambda_\alpha$ is a basis for say the adjoint representation of $\underline{H}$.

When $\chi^\alpha = \Lambda^\alpha \in C^\infty_0 (\mathbb{R}^3)$, we get the Gauss law operators 
$G(\Lambda)$. They vanish on physical quantum states.

The argument that local observables commute with $Q(\chi)$ is still valid so that the non-Abelian Sky 
group they generate is superselected. 

If $Q(\chi)|\cdot \rangle \neq 0$ when $\chi_\infty$ has a non-trivial angular dependence on $\hat{n}$, then 
this superselection sector is not Lorentz-invariant. Lorentz group is spontaneously broken.

All this is as before. But the operator $V(\omega)$ does not generalize to the non-Abelian case as its 
naive generalization does not commute with $G(\Lambda)$. Only $W(\theta)$ does so, with $S_{CS}(A)$ 
being the non-Abelian Chern-Simons term:
\begin{eqnarray}
W(\theta) &=& e^{i \theta S_{CS}(A)}, \\
S_{CS}(A)&=& \frac{1}{8\pi^2}{\rm Tr} \Big( A \wedge d A + \frac{2}{3} A \wedge A \wedge A \Big).
\end{eqnarray}
This operator commutes with $G(\Lambda)$, but not with $Q(\chi)$ for a generic $\chi$:
\begin{equation}
U(\chi) W(\theta) = C(\chi_\infty, \theta,A) W(\theta) U(\chi), \quad U(\chi) = e^{i Q(\chi)}, 
\end{equation}
where the explicit expression for $C(\chi_\infty, \theta,A)$ can be read off from Eq. (13.59) of \cite{Nair:2005iw}. It simplifies if $\chi^\alpha$ is 
non-zero for only one value $\alpha_0$ of $\alpha$. Any gauge transformation induced by $e^{i Q(\chi)}$ is 
a product of such transformations.

For the mass-deformed model, the deformed mass is a generalization of (\ref{csmass}) 
\begin{equation}
M(\theta) = m \cos \Big(\int \Re C(\chi_\infty, \theta, A)\Big).
\label{nabmass}
\end{equation}
It comes from absorbing the twist $W(\theta)$ of the twisted quantum state
\begin{equation}
|\cdot \rangle_\theta = W(\theta) |\cdot \rangle_0
\end{equation}
in the mass term.

The twisted sector of QCD with twist $\theta$  is precisely the QCD $\theta$-sector.It is the one where 
quarks and baryons acquire electric dipole moment. In the absence of Sky, there is $2\pi$-periodicity on 
$\theta$. But that is absent in $M(\theta)$: Lorentz violation is not periodic in $\theta$. 

In addition to modifying $m \bar{\Psi}\Psi$ by $U(\chi)$, we can also modify it by another operator discussed 
in \cite{Balachandran:2012bn}. That will let us calculate electric dipole moment as well. This operator too 
comes from a 
gauge principle and has shared properties with $U(\chi)$. They occur together in the twisted mass and 
mutually affect each other.

\section{Final Remarks}

There are particular features of {\it non-Abelian} gauge groups which merit more study. 

Let us focus on the subgroup of the non-Abelian Sky where $\chi_\infty$ is a constant function. We can write 
\begin{equation}
\chi_\infty(\hat{n}) = \chi_\infty^\alpha \lambda_\alpha, \quad \chi_\infty^\alpha \,\,{\rm constants}.
\end{equation}
Correspondingly,
\begin{equation}
Q(\chi) = \chi^\alpha_\infty Q(\lambda_\alpha) := \chi^\alpha_\infty Q_\alpha
\end{equation}
where $Q_\alpha$ generate the non-Abelian global group $H$. (Only the asymptotic value of $\chi$ 
matters because of the Gauss law.)

In QCD, $H$ is the color group $SU(3)_c$. It is what is normally studied in literature. So we can concentrate 
on $H$ and assume without loss of generality that $H$ is a simple compact Lie group. But $H$ is 
superselected. So are $Q_\alpha$. $H$ is also by assumption non-Abelian. Therefore not all elements 
$h = e^{i \phi_\alpha Q_\alpha}$ of $H$ can characterize a superselection sector. 

That is because local observables commute with $H$. So the vector states with the same value of a fixed 
$h_0 = e^{i \phi^0_\alpha Q_\alpha}$ are invariant under local observables. This value is part of the 
characterization of a superselection sector. 

But a generic $h \in H$, $H$ being non-Abelian, changes this value and hence the superselection sector. 
Such $h$ changing this value are "spontaneously broken". The choice of a particular value of $h_0$ is 
like choosing a direction for the Higgs field. 

Thus unlike the abelian $U(1)$ of electric charge or its group algebra $\mathbb{C}U(1)$, the full color 
group $SU(3)$ or its group algebra $\mathbb{C}SU(3)$ for instance cannot be used or implemented in 
a superselection sector. Global color cannot participate as a symmetry group in QCD.\\

\noindent {\it Remarks}\\

The above features of non-Abelian gauge groups were first encountered in the problem of color 
\cite{Balachandran:1983xz, Balachandran:1983fg}. 
It was subsequently understood as a generic feature of non-Abelian twisted bundles (see 
\cite{Balachandran:1991zj,Balachandran:1991ea}). It is encountered in numerous physical systems: 
molecules, non-Abelian monopoles, QCD, ... including when $H$ is a discrete group as for molecules 
\cite{Balachandran:1991ea}.

But we can use a maximal abelian subalgebra ${\cal A}_c$ of $\mathbb{C}H$ ($c$ for commutative) for 
labeling the superselection sectors.. If $H$ is a simple Lie group and $T^k$ is the maximal torus generated 
by its Cartan subalgebra, then it can be the algebra generated by the center $\mathcal{C}(\mathbb{C}H)$ of 
$H$ and the algebra $\mathbb{C}T^k$ of $T^k$. Thus ${\cal A}_c$ contains the Casimir invariant. 

Now ${\cal A}_c$ is {\it commutative}, so its irreducible representations are one-dimensional. 
Elsewhere \cite{Balachandran:2013kia} we have discussed this issue further. Here we summarize the main points.

If the $H$-bundle is trivial, we can show that pure states exist on the local algebra ${\cal A}$, but they 
are obtained by tracing out the $H$-degrees of freedom as ${\cal A}$ is $H$-blind. But if the $H$-bundle 
is twisted, the states obtained from a superselection sector and a GNS construction are all mixed. We will 
also show that the QCD bundle is twisted. 

Similar reasoning can be applied to quantum gravity. Previous work \cite{Balachandran:2011gj} 
treated mapping class groups partially from this perspective. Also in asymptotically flat spacetimes, the 
Poincar\'e group emerges as the analogue of the global gauge groups above in the ADM formalism. It is 
{\it non-Abelian} so that it is {\it spontaneously broken} to the analogue of ${\cal A}_c$. What this means 
requires more work to understand adequately. But it does indicate that a quantum black hole has features different from its classical description. 

For example, suppose we decide on $J_3$, the third component of angular momentum, as one unbroken 
generator creating an unbroken circle group in ${\cal A}_c$. But then all the $\hat{n} \cdot \vec{J}, \hat{n} 
\neq (0,0,1)$ are spontaneously broken, and $J_3$ has a fixed numerical value in  a superselection sector. 
No observable internal to the black hole can change it. It is not then clear if it makes quantum sense to 
interrogate if the black hole is spinning in a direction distinct from the $3$-direction, and with what angular 
velocity. 

The role of cluster decomposition, so important to analyze such issues for electric charge 
\cite{Haag:1992hx} has not been analyzed in quantum gravity. We remark that in QED, electric charge 
density $j_0$ is gauge invariant (commutes with $Q(\chi)$) and can be turned into a local observable. So 
there is quantum meaning for claims of measurement of charge in a finite volume $V$: it is the 
measurement of 
\begin{equation}
\int_V d^3 x \,j_0,
\end{equation}
a local observable. But for non-Abelian $H$, $Q_\alpha$ are {\it not} integrals of fully gauge invariant 
local densities $q_\alpha$ ($q_\alpha$ do not commute with all $Q(\chi)$). The same goes for say $J_3$ 
in quantum gravity: such observables are not integrals of local diffeomorphism-invariant densities. Thus 
while local measurements of electric charge density are possible, that is not so for non-Abelian $H$-charge 
densities or their diffeomorphism analogues in gravity. That already poses problems for us to adapt Haag's 
ideas (``positron behind the moon") to ${\cal A}_c$ or its gravity analogue, even though this algebra is abelian.\\

\noindent {\it Remark on Covariantization}\\

In the text, we have used the fixed-time formalism all the way. That leads to a ``blow-up" of spatial infinity 
where a sphere $S^2_\infty$ gets attached. 

This is enough to treat all but Lorentz boosts. But Lorentz boosts cannot act on such an infinity. We 
can accommodate the latter by attaching the $2+1$ de Sitter space, which carries the action of the 
Lorentz group, at spatial infinity. That this must be so can be inferred from the analysis of the Spi group 
\cite{Ashtekar:1978zz}. 

But for a satisfactory treatment, we need the covariant Peierls form of commutation relations 
\cite{Peierls:1952cb,DeWitt:2003pm,Bimonte:2003cx}. 
For free abelian electromagnetism, this can be written down compatibly with gauge invariance. For non-Abelian 
gauge theories including QCD, gauge invariance requires us to deal with an interacting non-linear theory.

But since the connected component of the Lorentz group ${\cal L}_+^\uparrow$ is simple, if rotations 
are spontaneously broken, so is ${\cal L}_+^\uparrow$. Thus indirectly, we can infer the fate of 
${\cal L}_+^\uparrow$.

We may finally note that the analysis in this paper can be readily adapted to all spacetimes of dimension 
$\geq2+1$. The $2+1$ case has in fact already been treated in \cite{Balachandran:2012md}.

\section{Acknowledgments}
APB was supported by the Institute of Mathematical Sciences, Chennai and by Instituto Internacional de 
Fisica, Natal, Brasil.

\end{document}